\documentstyle[psfig,twocolumn,aps,prl,epsf]{revtex}

\begin{document}

\title{Nuclear transparency from quasielastic A(e,e$'$p) reactions 
up to $Q^2$ = 8.1 (GeV/c)$^2$.}

\author {
K.~Garrow,$^{18}$
D.~McKee,$^{13}$
A.~Ahmidouch,$^{14}$ 
C.~S.~Armstrong,$^{18}$ 
J.~Arrington,$^{2}$
R.~Asaturyan,$^{22}$
S.~Avery,$^{7}$
O.~K.~Baker,$^{7,18}$
D.~H.~Beck,$^{8}$
H.~P.~Blok,$^{20}$
C.~W.~Bochna,$^{8}$
W.~Boeglin,$^{4,18}$
P.~Bosted,$^{1,a}$
M.~Bouwhuis,$^{8}$
H.~Breuer,$^{9}$
D.~S.~Brown,$^{9}$
A.~Bruell,$^{10}$
R.~D.~Carlini,$^{18}$
N.~S.~Chant,$^{9}$
A.~Cochran,$^{7}$
L.~Cole,$^{7}$
S.~Danagoulian,$^{14}$ 
D.~B.~Day,$^{19}$
J.~Dunne,$^{12}$
D.~Dutta,$^{10}$ 
R.~Ent,$^{18}$
H.~C.~Fenker,$^{18}$
B.~Fox,$^{3}$
L.~Gan,$^{7}$
D.~Gaskell,$^{2,16}$
A.~Gasparian,$^{7}$
H.~Gao,$^{10}$
D.~F.~Geesaman,$^{2}$
R.~Gilman,$^{17,18}$
P.~L.~J.~Gu\`eye,$^{7}$
M.~Harvey,$^{7}$
R.~J.~Holt,$^{8,b}$
X.~Jiang,$^{17}$
C.~E.~Keppel,$^{7,18}$ 
E.~Kinney,$^{3}$
Y.~Liang,$^{1,7}$
W.~Lorenzon,$^{11}$ 
A.~Lung,$^{18}$ 
D.~J.~Mack,$^{18}$
P.~Markowitz,$^{4,18}$
J.~W.~Martin,$^{10}$
K.~McIlhany,$^{10}$
D.~Meekins,$^{5,c}$ 
M.~A.~Miller,$^{8}$ 
R.~G.~Milner,$^{10}$ 
J.~H.~Mitchell,$^{18}$
H.~Mkrtchyan,$^{22}$
B.~A.~Mueller,$^{2}$
A.~Nathan,$^{8}$
G.~Niculescu,$^{15}$ 
I.~Niculescu,$^{6}$ 
T.~G.~O'Neill,$^{2}$
V.~Papavassiliou,$^{13,18}$
S.~Pate,$^{13,18m}$
R.~B.~Piercey,$^{12}$ 
D.~Potterveld,$^{2}$
R.~D.~Ransome,$^{17}$
J.~Reinhold,$^{4,18}$
E.~Rollinde,$^{18,21}$
P.~Roos,$^{9}$
A.~J.~Sarty,$^{5,d}$
R.~Sawafta,$^{14}$
E.~C.~Schulte,$^{8}$
E.~Segbefia,$^{7}$
C.~Smith,$^{19}$
S.~Stepanyan,$^{22}$
S.~Strauch,$^{17}$
V.~Tadevosyan,$^{22}$
L.~Tang,$^{7,18}$
R.~Tieulent,$^{9,18}$
A.~Uzzle,$^{7}$
W.~F.~Vulcan,$^{18}$
S.~A.~Wood,$^{18}$
F.~Xiong,$^{10}$
L.~Yuan,$^{7}$
M.~Zeier,$^{19}$
B.~Zihlmann,$^{19}$
and V.~Ziskin$^{10}$.
}

\address{
$^{1}${American University, Washington, D.C. 20016} \\
$^{2}${Argonne National Laboratory, Argonne, Illinois 60439} \\
$^{3}${University of Colorado, Boulder, Colorado 80309} \\
$^{4}${Florida International University, University Park, Florida 33199} \\
$^{5}${Florida State University, Tallahassee, Florida 32306} \\
$^{6}${The George Washington University, Washington, D.C. 20052} \\
$^{7}${Hampton University, Hampton, Virginia 23668} \\
$^{8}${University of Illinois, Champaign-Urbana, Illinois 61801} \\
$^{9}${University of Maryland, College Park, Maryland 20742} \\
$^{10}${Massachusetts Institute of Technology, Cambridge, Massachusetts 02139} \\
$^{11}${University of Michigan, Ann Arbor, Michigan 48109} \\
$^{12}${Mississippi State University, Mississippi State, Mississippi 39762} \\
$^{13}${New Mexico State University, Las Cruces, New Mexico 88003} \\
$^{14}${North Carolina A \& T State University, Greensboro, North Carolina 27411} \\
$^{15}${Ohio University, Athens, Ohio 45071} \\ 
$^{16}${Oregon State University, Corvallis, Oregon 97331} \\
$^{17}${Rutgers University, New Brunswick, New Jersey 08903} \\
$^{18}${Thomas Jefferson National Accelerator Facility, Newport News, Virginia 23606} \\
$^{19}${University of Virginia, Charlottesville, Virginia 22901} \\
$^{20}${Vrije Universiteit, 1081 HV Amsterdam, The Netherlands} \\ 
$^{21}${College of William and Mary, Williamsburg, Virginia 23187} \\
$^{22}${Yerevan Physics Institute, 375036, Yerevan, Armenia} \\
}

\date{\today}
\maketitle

\begin{abstract}
The quasielastic (e,e$^\prime$p) reaction was studied on targets of 
deuterium, carbon, and iron up to a value of momentum 
transfer $Q^2$ of 8.1 (GeV/c)$^2$. A nuclear transparency was determined
by comparing the data to calculations in the Plane-Wave Impulse Approximation.
The dependence of the nuclear transparency on $Q^2$ and the mass number $A$ was
investigated in a search for the onset of the Color Transparency phenomenon. 
We find no evidence for the onset of Color Transparency within
our range of $Q^2$. A fit to the world's nuclear transparency data
reflects the energy dependence of the free proton-nucleon cross section.
\end{abstract}

\section{Introduction}

The concept of Color Transparency (CT) was introduced two decades ago
by Mueller and Brodsky \cite{CT}, and since has stimulated great experimental
and theoretical interest. CT is an effect of QCD, related to the presence of
non-abelian color degrees of freedom underlying strongly interacting matter. 
CT has its most unique manifestation in A(p,2p) or A(e,e$^\prime$p) experiments
at high energies. The basic idea is that, under the right conditions,
three quarks, each of which would normally interact very strongly with
nuclear matter, could form an object that passes undisturbed through the nuclear
medium. A similar phenomenon occurs in
QED, where an $e^+e^-$ pair of small size has a small cross section
determined by its electric dipole moment \cite{perkins}. In QCD, a 
$q \bar{q}$ or $qqq$ system can act as an analogous small color dipole moment.

CT was first discussed in the context of perturbative QCD. 
Later work \cite{CTreview} indicates that this phenomenon also occurs in a
wide variety of model calculations with non-perturbative reaction mechanisms.
In general, the existence of CT requires that high momentum transfer
scattering takes place via selection of amplitudes in the initial and 
final state hadrons characterized by a small transverse size.  Secondly, this
small object should be `color neutral' outside of this small radius in order
not to radiate gluons.  Finally, this compact size must be maintained for
some distance in traversing the nuclear medium.
Unambiguous observation of CT would provide a new means to
study the strong interaction in nuclei.

Several measurements of the transparency of the nuclear medium to
high energy protons in quasielastic A(p,2p) and  A(e,e$^\prime$p)
reactions have been carried out over the last decade.
The nuclear transparency measured in A(p,2p) at Brookhaven \cite{bnl}
has shown a rise consistent with CT for Q$^2 \simeq$ 3 - 8 (GeV/c)$^2$,
but decreases at higher momentum transfer.
At the time, questions were raised about the interpretation of
this data, as only one of two final-state protons was momentum-analyzed,
and the exclusivity of the reaction could not be guaranteed.
A more recent experiment \cite{bnlnew}, completely reconstructing
the final-state of the A(p,2p) reaction, confirms the validity
of the earlier Brookhaven experiment.
Two explanations for the surprising behavior were given: 
Ralston and Pire \cite{ralpire}
proposed that the interference between short and long distance amplitudes
in the free $p$-$p$ cross section was responsible for these energy
oscillations, where the nuclear medium acts as a filter for the
long distance amplitudes. Brodsky and De Teramond
\cite{brodsky} argued that the unexpected
decrease could be related to the crossing of the open-charm threshold.

The NE-18 A(e,e$^\prime$p) measurements at SLAC \cite{ne18q,ne18a} yielded
distributions in missing energy and momentum completely consistent with
conventional nuclear physics predictions. The extracted transparencies
exclude sizable CT effects up to $Q^2$ = 6.8 (GeV/c)$^2$, in contrast to
the A(p,2p) results \cite{bnl}. The measurements ruled out several models
predicting an early, rapid, onset of CT, but could not exclude models
predicting a slow onset of CT. The proposed explanation of Ralston and Pire
\cite{ralpire}, that the nuclear medium $A$ eliminates the long distance
amplitudes in the A(p,2p) case, might resolve the apparent discrepancy
between the A(e,e$^\prime$p) and A(p,2p) results. Still, questions remain with
the recent claim that the nuclear transparencies at $Q^2 \simeq$ 8 (GeV/c)$^2$
in A(p,2p) experiments deviate from Glauber predictions \cite{bnlnew}.

Intuitively, one expects an earlier onset of CT for meson production
than for hard proton scattering, as it is much more probable to produce a
small transverse size in a $q \bar{q}$ system than in a three quark system.
By contrast, microscopic calculations for meson production from nuclei may
be on less solid footing than in the comparable A(e,e$^\prime$p) case.
Nuclear transparencies in exclusive incoherent $\rho^0$ meson production from
nuclei have been measured in several experiments.
At Fermilab \cite{adams}, increases in
the nuclear transparencies have been observed as the virtuality of the photon
increases, as expected from CT. Inclusion of CERN data on similar nuclear
transparencies, at higher $Q^2$, however, make the effect less significant
\cite{arneo}. In addition, for such reactions one has to distinguish
coherence length from formation length effects. The coherence length is
the distance at which the virtual photon fluctuates
into a $q\bar{q}$ pair. The formation length is the
distance traveled by the small size $q\bar{q}$ pair before evolving
to the normal $\rho$ meson size. Formation lengths are the
governing scale to look for CT effects. Evidence for strong
coherence length effects has recently been reported by the HERMES
experiment at DESY \cite{hermes}. It should be noted that
for the Fermilab experiment formation lengths
are only a factor of approximately 2-3 larger than coherence lengths.

Recent support for CT comes from the coherent diffractive 
dissociation of 500~GeV/c negative pions into di-jets \cite{fnal}.
Such a di-jet production reaction is not an exclusive reaction, and may
thus differ fundamentally from other searches for CT.
The inferred $Q^2$ for this reaction was larger than 7~(GeV/c)$^2$.
The A-dependence of the data was fit assuming $\sigma \propto A^{\alpha}$
for three $k_t$ bins, with $k_t$ the jet transverse momentum.
For $1.25 < k_t < 1.5$ GeV/c, $1.5 < k_t < 2.0$ GeV/c, and $2.0 < k_t < 2.5$ GeV/c
($Q^2 \geq 4k_t^2$)
the alpha values were determined to be $\alpha$=1.64$^{+0.06}_{-0.12}$,
$\alpha$=1.52$\pm0.12$, and $\alpha$=1.55$\pm0.16$, respectively.
This is far larger than the $\sigma \propto A^{0.7}$ dependence typically
found in inclusive $\pi$-nucleus scattering, whereas the theoretical
\cite{frank93} CT values were predicted to be $\alpha$ = 1.25, 1.45
and 1.60, respectively. These di-jet data, however, do not inform about
the kinematic onset of CT. To date, none of the mentioned measurements
provides direct information on the onset of CT.

Quasielastic A(e,e$^\prime$p) reactions have several advantages to
offer in searching for CT effects. The fundamental $e$-$p$ scattering
cross section is smoothly varying and accurately known;
compared to the A(p,2p) reaction one has less sensitivity to the
unknown large momentum components of the nuclear wave function \cite{farrar};
energy resolutions are sufficient to guarantee the exclusivity of the reaction;
and, one does not have to distinguish coherence length effects.
The purpose of the present experiment was to measure the nuclear transparency
in the A(e,e$^\prime$p) reaction with greatly improved statistics and
systematic uncertainties compared to the NE-18 experiment \cite{ne18q,ne18a},
and to increase the $Q^2$ range in order to search for the onset of CT.
The precision of the presented data, in addition to the reliability of
conventional nuclear transparency calculations for the A(e,e$^\prime$p)
reaction, allows for a conclusive test of such an onset.


\section{Experiment and Data Analysis}

The experiment was performed in Hall C at the Thomas
Jefferson National Accelerator Facility (TJNAF). Beam energies of 3.059,
4.463, and 5.560 GeV were used for the $Q^2$ values of 3.3, 6.1, and 8.1
(GeV/c)$^2$, respectively. The electron beam impinged on either a cryogenic
target system, consisting of 4.5 cm long liquid hydrogen and deuterium targets,
cooled to 19 K and 22 K, respectively, or a solid target system, which
incorporated a solid $^{12}$C target of 3\% radiation length and solid
$^{56}$Fe targets of 3\% and 6\% radiation length. The target thickness
uncertainty is estimated to be 0.3\% for the solid targets, and 0.5\% for
the liquid targets.

An aluminum dummy target, consisting of two 0.99 mm Al targets separated
by 4.5 cm, was suspended from the cryogenic target system in order to
subtract the 0.13 mm Al window contributions to the cryogenic targets.
Beam currents ranged from 30 to 60~$\mu$A depending on the target used.
The beam current was measured by a system of beam current monitors along with
a parametric current transformer for absolute calibration. The error in the
absolute calibration due to noise was less than 0.2 $\mu$A. Thus, the error
in the accumulated beam charge is less than 1\%.
Hall C's High Momentum Spectrometer (HMS) and Short Orbit Spectrometer (SOS)
were used to detect the knocked-out proton and scattered electron,
respectively. The spectrometers and their detection packages are
described in Ref. \cite{donprl}.
The momentum acceptance ($\frac{\Delta p}{p}$) utilized in the HMS was
$\pm 8\%$, and in the SOS $\pm 15\%$. 
The various kinematics are given in Table I.

In addition to the coincident A(e,e$^\prime$p) data, 
for all kinematics a subset of single events was recorded
with a statistical accuracy of much better than 1$\%$. This enabled
monitoring the product of detector efficiencies, accumulated charge
and target density effects on a run-to-run basis.
Since the target density of the cryogenic targets is influenced
by heating effects due to the incident electron beam, a correction
was applied. This correction was
(2.0$\pm$0.4)\% for the highest beam current (60 $\mu$A).

Prior to the start of the A(e,e$^\prime$p) experiment,
$^1$H(e,e$^\prime$) elastic electron-proton events were recorded
in both spectrometers at a beam energy of 2.056 GeV.
Both spectrometers measured a fixed scattered
electron momentum of 1.350 GeV/c, while the spectrometer angles were
varied from 32.9$^\circ$ to 42.9$^\circ$
to scan the elastically scattered electrons across the spectrometer
momentum acceptance. The known hydrogen cross section \cite{bosted} was then
used to check the spectrometers for both normalization and acceptance
problems. The measured and simulated HMS data agreed to better
than 2\% for the entire momentum acceptance used in the data analysis.
The SOS acceptance, however, showed a complicated correlation
among the vertex position ($y_{target}$); the angle of the scattered electron 
($y'_{target}$); and the momentum deviation of the scattered electron
($\delta p/p$). The latter two are defined with respect to the
nominal spectrometer angle and momentum, respectively.
Simulations showed that such correlated effects become important when using
a target with an effective target length (i.e., the target length as viewed
by the spectrometer) larger than about 2.5 cm.
For elastic scattering $y'_{target}$ and $\delta p/p$ are correlated, and we
found normalization problems at large $\pm y_{target}$ and large
$\pm y'_{target}$. The H(e,e$^\prime$) calibration runs approached a
normalization consistent with the expectations based upon the world
H(e,e$^\prime$) cross sections \cite{bosted}, however,
when a small $y_{target}$ cut was used.

In the actual A(e,e$^\prime$p) data taking the SOS was positioned at far larger
scattering angles (see Table I), to obtain the highest possible $Q^2$, and was
thus even more susceptible to this acceptance problem.
Although we do believe that we have understood the acceptance problem,
we took the simple solution of normalizing the D(e,e$^\prime$p) data to
the H(e,e$^\prime$p) results. Any lack of understanding of the extended
target acceptance cancels in the ratio of yields of two similar,
extended targets.
It was, indeed, verified that the ratio of the deuterium and hydrogen yields,
within statistics, does not depend on the cut on $y_{target}$.
It should be noted that the measurements on
the almost point-like solid targets
are not affected by the mentioned acceptance problem.

For the A(e,e$^\prime$p) results,
the yields were corrected for proton absorption
in the target and through the various components of the spectrometer.
This correction varied from 5 to 6.5\% depending on the target used.
The correction could be partially checked by comparing elastic
$^1$H(e,e$^\prime$) and $^1$H(e,e$^\prime$p) rates.
The uncertainty in the correction is estimated to be 1\%.

Coincident detection of the recoil electron and ejected proton momentum
enabled the determination of the energy transfer, $\nu = E_e - E^{\prime}_e$,
where $E_e$ is the electron beam energy and $E^{\prime}_e$ is the 
energy of the detected electron, and the missing energy,
$E_m = \nu - T_{p^\prime} - T_{A-1}$,
where $T_{p^\prime}$ and $T_{A-1}$ are the kinetic energy of the final state proton
and $A-1$ recoil nucleus, respectively. Also, the missing momentum
$\vec{p}_m = {\vec p}_{p^\prime} - \vec{q}$, where ${\vec p}_{p^\prime}$
and $\vec{q}$ are the momentum of the detected proton and the
three momentum transfer in the interaction, can be computed. 
The missing energy $E_m$ is equal to the separation energy, $E_s$,
needed to remove the nucleon from a particular state within the nucleus. 
Assuming the plane-wave impulse approximation (PWIA) to be valid,
the missing momentum $\vec{p}_m$ is equal to the 
initial momentum of the proton within the nucleus.
In a non-relativistic PWIA formalism, the cross section can be written
in  a factorized form as
\begin{equation}
\label{sig_ep}
\frac{d^6 \sigma}{dE^{\prime}_e d\Omega^{\prime}_e dE_p 
d\Omega_p}
=  K\sigma_{ep}S(E_m,\vec{p}_m),
\end{equation}
where $dE^{\prime}_e, d\Omega^{\prime}_e, dE_p$ and
$d\Omega_p$ are the phase space factors of the electron and proton,
$K$ = $\vert {\vec p}_{p^\prime} \vert E_p$ is a known kinematical factor, 
and $\sigma_{ep}$ is the off-shell electron-proton cross section.
The choice of off-shell cross section \cite{deforest} is set by choosing
a prescription to apply momentum and energy conservation at the $\gamma_vp$
vertex. Here, $\gamma_v$
is the virtual photon with energy $\nu$ and 3-momentum $\vec{q}$, and
$p$ represents an off-shell proton, with initial momentum
$\vec{p_p}$ and separation energy $E_m$. 
The spectral function $S(E_m,\vec{p}_m)$ 
is defined as the joint probability of finding a proton of
momentum $\vec{p}_m$ and separation energy $E_m$ within the
nucleus. This function contains the nuclear structure information
for a given nucleus. 

The definition of the transparency ratio is the same
as in the early, pioneering A(e,e$^\prime$p) CT experiment \cite{ne18q,ne18a},
that is the ratio of the cross section measured in
a nuclear target to the cross section
for (e,e$^\prime$p) scattering in PWIA. Numerically this
ratio can be written as
\begin{equation}
\label{trans}
T(Q^2) = \frac{\int_V d^3p_m dE_m Y_{exp}(E_m,\vec{p}_m)}
{\int_V d^3p_m dE_m Y_{PWIA}(E_m,\vec{p}_m)},
\end{equation}
where the integral is over the phase space $V$
defined by the cuts $E_m <$~80~MeV and $\left|{\vec{p}_m}\right| <$~300~MeV/c,
$ Y_{exp}(E_m,\vec{p}_m)$ and $Y_{PWIA}(E_m,\vec{p}_m)$
are the corresponding experimental and simulation yields.
The $E_m$ cut prevents inelastic
contributions above pion production threshold.

The off-shell prescription $\sigma^{cc}_1$ of Ref.
\cite{deforest} was used for the evaluation of $\sigma_{ep}$ in
Eq. \ref{sig_ep}. The measured nuclear transparencies 
are hardly sensitive to the inclusion of such off-shell effects,
- using an on-shell form changes $T$ by less than 1\%.
The spectral functions used as input to the simulation
are the same as in Refs. \cite{ne18q,ne18a}. 
The distribution of events in $E_m$ (describing knockout from particular
orbits) is characterized by Lorentzian energy profiles to account for the
spreading width of the one-hole states.
The momentum distributions are calculated using a Woods-Saxon nuclear
potential with shell-dependent parameters. The Lorentzian and Woods-Saxon
parameters are determined from fits to spectral functions extracted from
previous A(e,e$^\prime$p) experiments \cite{FM84}.
Descriptions of the deepest-lying shells
of Fe were taken from a Hartree-Fock calculation \cite{Negele}
since data on these shells are inconclusive.

These spectral functions
are generated with the assumption of an independent-particle model,
which is known to overestimate the experimental yield from a given shell
model configuration, as defined via ($E_m$,$|\vec{p}_m|$) limits.
Nucleon-nucleon correlations are not contained in such a
formalism and are known to move independent-particle model yields to higher
excitation energies. In order to account for this a so-called correlation
correction was applied. The correlation corrections
for the kinematical cuts applied to the data, $E_m <$ 80 MeV and 
$\left|{\vec{p}_m}\right| <$ 300 MeV/c,
were 1.11 (1.22) for $^{12}$C ($^{56}$Fe); these corrections have
uncertainties estimated to be 0.03 (0.06). These correlation corrections
have been previously determined from $^{12}$C and ${}^{16}$O
\cite{vanorden} spectral functions that include the effects of correlations.
For Fe, a correlated nuclear matter spectral function corrected for finite
nucleus effects \cite{ji} was used to estimate this correction factor.
These correlation corrections would correspond to spectroscopic factors that
are higher than what has been determined from lower $Q^2$ A(e,e$^\prime$p) data,
by typically 20\% (or 1-2$\sigma$) \cite{louk,dutta}. This is an unresolved issue.
One can not determine spectroscopic factors independently from nuclear
transparencies. Here, we use the mentioned correlation corrections for
consistency with previous nuclear transparency data \cite{ne18q,ne18a,donprl}.

The measured $^{12}$C(e,e$^\prime$p) yields, as function of missing
momentum, and the predictions from the simulation are shown in Fig. 1. The
requirement that $E_m<$80~MeV was applied to both data and
Monte Carlo distributions. Good agreement between the
momentum distributions is observed for all $Q^2$ points measured.
For the $^{56}$Fe(e,e$^\prime$p) case good agreement is found
for the momentum distributions, but discrepancies between data and simulations
can be observed in the missing energy distributions.
 
For the lowest $Q^2$ point of 3.3~(GeV/c)$^2$ the Fermi cone
was mapped by varying the angle of the proton
spectrometer about the quasi-free angle. The upper
panel of Fig. 2 shows the normalized yield for the various 
angles about the quasi-free angle for the $^{12}$C and $^{56}$Fe targets.
The solid symbols represent the results of the present measurement
while the open symbols from a previous measurement \cite{donprl} 
are shown for comparison. The solid line in the top panel represents
the predictions for the Monte Carlo yield. The lower panel
of Fig. 2 shows the extracted transparency ratio for the 
$^{12}$C and $^{56}$Fe targets at the various angles measured.
The solid line represents the statistically
averaged transparency ratio from the
present results and the result is seen to coincide with the 
central value within the errors of the measurement. Again the data
(open symbols) from the previous measurement \cite{donprl} 
are shown for comparison. The transparency ratio is also seen
to be close to constant for the various angles about the
quasi-free angle. In the previous measurement \cite{donprl}
large asymmetries were seen in the ratios about the
quasi-free angle for $Q^2$=0.8 and 1.3~(GeV/c)$^2$. This was
interpreted as the presence of a longitudinal-transverse
interference term in the measured cross section approximately
20$\%$ larger than what is contained in the off-shell 
$\sigma_{ep}$ cross section.
The disappearance of this asymmetry indicates that
the reaction mechanism is simpler at these larger values of $Q^2$.

The nuclear transparencies for deuterium, carbon and iron
are given in Table II for the various $Q^2$ values measured.
Typically, the point-to-point systematic uncertainty amounts to 2.3$\%$,
dominated by uncertainties in current measurement (0.7\%), run-to-run
stability of (e,e$^\prime$) and (e,p) singles events ($<$1\%), and an
estimated 1\% for the proton absorption correction applied.
The quoted 2.3\% does neither take into account a normalization-type
uncertainty of 3\%, nor the model-dependent
systematic uncertainties implicit in the extraction of the transparency ratios.
The normalization-type uncertainty is mainly due to the radiative corrections,
the choice of electron-proton cross section, and acceptance.
The model-dependent uncertainties are target nucleus dependent and are due to
choices in
spectral function parameters and the uncertainty in correlation correction.

There are some exceptions to the 2.3\% point-to-point systematic uncertainty.
As mentioned previously,
the deuterium transparency results were obtained by dividing
by the corresponding measured hydrogen cross section data. This
accounts for the normalization problems in the
deuterium target due to the effects of the extended target.
The results then were in good agreement with the earlier measurement
\cite{ne18q,ne18a}. Nonetheless, a larger systematic uncertainty of 3\%
was assigned to the deuterium results. The iron measurement at
$Q^2$=6.1~(GeV/c)$^2$ also was assigned a larger systematic uncertainty
of 3.8\%, because of uncertainty due to target thickness.


\section{A-Dependence}

The measured transparency $T(Q^2)$ values from this (large solid symbols)
and previous work are presented in Fig. 3. The errors shown include
statistical and systematic uncertainties, but do not include model-dependent
systematic uncertainties in the spectral functions and correlation corrections
used in the simulations. This is the same as for the
data of Ref. \cite{donprl} (small solid symbols).
Data from previous experiments \cite{ne18q,ne18a,bates} (represented by open
symbols) include the full uncertainty. For completeness, we also show results
using gold targets, from previous experiments only. The present results 
for carbon and iron are of similarly high precision as those of 
Ref. \cite{donprl},
and of substantially higher precision than of Refs. \cite{ne18q,ne18a,bates}.
Our results at $Q^2$ = 3.3 (GeV/c)$^2$ agree well with previous results
for deuterium \cite{ne18q,ne18a}, carbon, and iron \cite{donprl}.

Little or no $Q^2$-dependence can be seen in the nuclear transparency data
above $Q^2 \approx$ 2 (GeV/c)$^2$. Excellent constant-value fits can be
obtained for the various transparency results above such $Q^2$. For
deuterium, carbon, and iron fit values are obtained of
0.904 ($\pm$ 0.013), 0.570 ($\pm$ 0.008), and 0.403 ($\pm$ 0.008),
with $\chi^2$ per degree of freedom of 0.56, 1.29, and 1.17, respectively.
As in Ref. \cite{donprl}, we compare with the results from correlated Glauber
calculations, including rescattering through third order \cite{glauber},
depicted as the solid curves for 0.2 $< Q^2 <$ 8.5 (GeV/c)$^2$.
In the case of deuterium, we show (dashed curve) a generalized Eikonal
approximation calculation, coinciding with a Glauber approximation for
small missing momenta \cite{sargsian}.
The $Q^2$-dependence of the nuclear transparencies is well described, but
the transparencies are underpredicted for the heavier nuclei. This behavior
persists even taking into account the model-dependent systematic
uncertainties.

Recently, a new calculation of nuclear transparencies has become available
\cite{zhalov}. This results in a better agreement between Glauber calculations
and the $A$-dependence of the nuclear transparency data. In this paper
\cite{zhalov} it was argued that the uncertainty in the treatment of
short-range correlations in the Glauber calculation can be constrained with
inclusive A(e,e$^\prime$) data.
This results in an effective renormalization of the nuclear
transparencies for the $^{12}$C and $^{56}$Fe nucleus of 1.020
and 0.896, respectively. Such a renormalization is due to integration
of the denominator in Eq. 2 over a four-dimensional phase space $V$
in $E_m$ and $\left|{\vec{p}_m}\right|$ argued to be more consistent with
experiments. I.e., the experiment measures an angular distribution in the
scattering plane rather than the complete
$\left|{\vec{p}_m}\right| <$ 300 MeV/c region.
This reduces the influence of short-range correlations. The nuclear
transparencies as given in Table II would have to be multiplied by these
renormalization factors, rendering values more consistent with the $A$-dependence
of Glauber calculations.
Although such a renormalization may be appropriate, we
quote nuclear transparency numbers consistent with the procedure of Refs.
\cite{ne18q,ne18a,donprl}, for the sake of comparison.

For the remainder of this Section, we will concentrate on a combined
analysis of the world's A(e,e$^\prime$p) nuclear transparency data.
Figure 4 shows $T$ as a function of $A$. The curves represent empirical fits
of the form $T = cA^{\alpha(Q^2)}$, using deuterium, carbon, and iron data.
We find, within uncertainties, the constant $c$ to be consistent with unity
as expected and the constant $\alpha$ to exhibit no $Q^2$ dependence
up to $Q^2$ = 8.1 (GeV/c)$^2$.
A similar treatment to nuclear transparency results of the older
A(e,e$^\prime$p) experiments renders a nearly constant value of
$\alpha$ = $-$0.24 for $Q^2 \geq$ 1.8 (GeV/c)$^2$. Numerical values are
presented in Table III. We note that using the renormalizations of the
nuclear transparencies proposed by Frankfurt, Strikman, and Zhalov
\cite{zhalov} would
reduce the numerical values of $\alpha$ by approximately 0.03.

Alternatively, we can analyze the $T$ results from the different nuclei
($A \geq$ 12), and the different experiments, in terms of a simple
geometric model, similar to that used in Refs. \cite{ne18a,sigeff}.
This model assumes classical
attenuation of protons propagating in the nucleus, with an effective
proton-nucleon cross section $\sigma_{\rm eff}$ that is independent of density:
\begin{equation}
T_{\rm class} = {1 \over Z}\int d^3r \, \rho_Z
({\bf r})\exp{ [-\int dz' \, \sigma_{\rm eff} \rho_{A-1} ({\bf  
r'})]}\;.
\label{eq:tclass}
\end{equation}
For this calculation, the nuclear (charge) density distributions were taken from
Ref. \cite{deVries} and $\sigma_{\rm eff}$ is the only free parameter.
The difference in number of protons and neutrons for a heavy nucleus is taken
into account in constructing $\sigma_{\rm eff}$. It is possible
that the effective change of $\sigma_{pp}$ in a nuclear medium is different
from that of $\sigma_{pn}$, but this is neglected.
Finally, we also assume
that the hard scattering rate is accurately determined by our PWIA model.
Therefore, any energy dependence of the transparency is due to Final-State
Interactions. Note that in the limit of complete CT, one would expect
$\sigma_{\rm eff} \rightarrow 0$.
The results of fitting this model to the measured $T$ results are shown
in Table III and Fig. 5. We also show, both in Table III and Fig. 5, the
results of fits using the $T$ values of Refs. \cite{ne18q,ne18a,donprl,bates}.
Using the renormalization factors of the nuclear transparencies advocated
by \cite{zhalov} results in values for $\sigma_{\rm eff}$ which are
reduced by approximately 2-3 mb.

We compare, in Fig. 5, the results for $\sigma_{\rm eff}$ with a 
normalized parameterization of the free proton-nucleon scattering
total cross sections \cite{pdb} (solid curve). We observe no
noticable energy dependence of $\sigma_{\rm eff}$ beyond that
of the free proton-nucleon scattering data. Thus, most of the variation of
$T(A)$ as a function of $Q^2$ is a reflection of the energy-dependence
of the free $N-N$ total cross section. In free proton-nucleon scattering, the
minimum at $T_{p^\prime} \approx$ 500 MeV is especially prominent \cite{pdb}, 
affecting the $T(A)$ values at $Q^2 \leq$ 1.3 (GeV/c)$^2$.
 
Regarding the normalization, we
find, with a $\chi^2$ per degree of freedom of 0.9,
an effective proton-nucleon cross section reduced by (71.4 $\pm$ 2.4)\% 
with respect to the free proton-nucleon cross section. 
Such a reduction has been interpreted \cite{ne18a} as
effectively taking into account effects such as Pauli blocking 
(at low energies) and short-range correlations,
but is mainly an artifact of the simple geometric model used in Eq. 3.
E.g., if one would fit, within the Glauber calculations of Ref. \cite{frankfurt},
an effective proton-nucleon cross sections to the present $^{12}$C(e,e$^\prime$p)
nuclear transparency data, one would find far closer agreement with
the free proton-nucleon total cross sections. On average, the reduction one
finds in such a procedure is less than 10\%.

Naively, the near-constancy of the effective proton-nucleon cross section as function
of $Q^2$, up to $Q^2$ = 8.1 (GeV/c)$^2$, seems to rule out the onset of CT.
However, the near-constancy of transparencies versus $Q^2$, as shown in Fig. 3,
may also result from a cancelling of effects in the hard electron-proton
scattering and CT effects in the nucleon propagation \cite{sigeff}.
One could argue that medium-dependent effects on hard electron-proton
scattering will have a different $A$- and $Q^2$-dependence than CT effects,
but the geometric model used here is obviously too simple to incorporate a
full description of the $A$-dependence of the data. 

If CT effects can be ruled out within the kinematics of the reported
data, the near-constancy of both the effective proton-nucleon cross
section and the nuclear transparencies as function of $Q^2$ suggests that the
quasi-free electron-proton scattering cross section equals the free
electron-proton scattering cross section (corrected for off-shell
effects as in Ref. \cite{deforest}).
If interpreted as constraining the medium-modification
of the proton magnetic form factor $G_M(Q^2)$, the $T$ results for
$Q^2 \geq$ 1.8 (GeV/c)$^2$ rule out a larger than 3\% variation in
the magnetic charge radius. 

The typical effective proton-nucleon cross section found from this data
analysis is approximately 30 mb. This is much larger than that derived from
the A(p,2p) data of Ref. \cite{bnl}, translated to similar values of $Q^2$ or
$T_{p^\prime}$. Jain and Ralston \cite{sigeff} derive values of approximately
15 mb from the latter data, using the same geometric model.
It seems that a discrepancy exists, likely just related to both the validity
of the simple geometric model used and the validity of the concept of an
effective proton-nucleon cross section.

As mentioned earlier,
the recent A(p,2p) data \cite{bnlnew} confirm the earlier trend in nuclear
transparency. The data agree reasonably well with Glauber calculations 
\cite{frankfurt} at
incident beam momenta of 6 GeV/c and about 12 GeV/c, and show a rise and
subsequent decrease in nuclear transparency in between. This rise and
decrease seem consistent with the ratio of observed $p-p$ cross
section and the predicted hard scaling behavior. Thus, the nuclear filtering
as proposed by Ralston and Pire \cite{ralpire} may be responsible for the
apparent contradiction between the proton transparency results from
A(p,2p) and A(e,e$^\prime$p) results, in similar regions of $Q^2$.
If so, it is not clear why the A(p,2p) data numerically agree with
Glauber calculations at incident beam momenta of 6 and 12 GeV/c,
but not at 9 GeV/c. Alternatively, the apparent discrepancy may be related to the
observation that the sensitivity to large momentum components in the
nuclear wave function is different for A(p,2p) and A(e,e$^\prime$p)
\cite{farrar}. Regardless, it seems that a $Q^2$ of 8 (GeV/c)$^2$
is not sufficient yet to select small transverse size objects in the
hard $e-p$ scattering process.

\section{$Q^2$-Dependence}

The $^{12}$C(e,e$^\prime$p) reaction has important benefits for a detailed
study of the onset of CT. The $^{12}$C nucleus has a relatively simple nuclear
structure, and previous low-$Q^2$ measurements provide accurate
information on its spectral function.
Quasi-free electron-proton scattering rates off $^{12}$C nuclei are large due
to the reduced transparency effects with respect to heavier target nuclei,
which, although it reduces the sensitivity to CT effects, enables the use of
statistics to perform studies of systematic uncertainties. 
In addition, a $^{12}$C target can, unlike e.g. an $^{56}$Fe
target, thermally withstand high (100 $\mu$A) electron beam currents.
Thus, both the previous TJNAF experiment
\cite{donprl} and the present experiment obtained results of high precision
in both statistics and systematics for nuclear transparencies determined from
the $^{12}$C(e,e$^\prime$p) reaction.
For these reasons, we used the $^{12}$C(e,e$^\prime$p) results to perform
a statistical analysis of the $Q^2$ dependence of the nuclear transparency.

The $^{12}$C(e,e$^\prime$p) nuclear transparency results are shown in Fig. 6,
with several state-of-the-art calculations that do or do not include
CT effects \cite{glauber,frankfurt,nikolaev,ralston}. To reduce the influence
of the energy-dependence of the $N-N$ total cross section, we restricted the
analysis to energies substantially above the minimum in this cross section,
i.e. to $Q^2$ values above  1.8 (GeV/c)$^2$.
Additionally, the normalizations of the various calculations were, in the
statistical analysis, treated as a free parameter,
as approximations concerning e.g. the influence of
short-range correlations and the density-dependence of the $N-N$ cross section
will affect the
absolute magnitude of the nuclear transparencies calculated (but have little
influence on the $Q^2$ dependence for large enough $Q^2$).
This enhances the sensitivity to a possible $Q^2$ dependence predicted by
the inclusion of CT effects.

As in Fig. 3, the data are once more compared in Fig.~6 with the results of the
correlated Glauber calculation of Ref. \cite{glauber} (solid curve).
In addition, various other calculations are shown.
Kundu {\it et al.} \cite{ralston} follow a perturbative QCD approach in the
Impulse Approximation. Due to the hard scattering, only the short distance
distribution amplitudes dominate. The ``expansion'' or diffusion in the quantum
mechanical propagation of quarks sideways and longitudinally is included in the
perturbative treatment. The effects of interaction with the nuclear medium
is included through an Eikonal form. The calculation has to make an
assumption on the distribution amplitudes. It appears \cite{ralston} that
perturbative QCD effects are better applied to the nuclear medium, due to
suppression of long distance components, and that CT effects are slower
for end point-dominated (``double-bump'') distribution amplitudes,
of e.g. Refs. \cite{cz,ks}, rather than for the asymptotic (``single-bump'')
distribution amplitude. Here, a calculation with the distribution
amplitude of Ref. \cite{ks} is shown (dashed curve).
A calculation of the effective proton-nucleon cross section of Ref.
\cite{ralston}, within the same framework, is added as a dot-dashed curve in
Fig. 5, and is almost coinciding with our fit result. This may just
reflect a similar neglect of detailed nuclear physics effects as in
the simple geometric model of Eq. 3.

By contrast, Ref. \cite{frankfurt} uses a more classical Distorted-Wave Impulse
Approximation approach, starting from a realistic ansatz for the nuclear
structure wave function and the optical limit to incorporate distortion
effects. A quantum diffusion model is used to describe the expansion of
the small size configuration selected in the (hard) scattering process
to its physical size. The model depends on the hadronization length as
a parameter, which in turn depends on the mass difference squared,
$\Delta M^2$, between
the proton and the first inelastic diffractive intermediate state.
The thick solid curve represents the correlated Glauber calculation of Ref.
\cite{frankfurt}. The dot-dashed curve represents the calculation
with the inclusion of CT effects, under the assumption that $\Delta M^2$
= 1.1 GeV$^2$ \cite{frankfurt}.

Lastly, Nikolaev {\it et al.}
\cite{nikolaev} assume closure is allowed within the reasonably
broad $E_m$ and $p_m$ acceptance of the nuclear transparency experiments,
and calculates Final-State Interaction effects in the Glauber approximation.
The calculation argues against an assumed factorization
into a PWIA model and a global attenuation factor. CT effects, due to the
interference of the elastic and inelastic intermediate states, are included
based on an expansion of the struck nucleon wave function in terms of
excited hadronic basis states. The dotted curve represents the result of
this calculation.

Table IV displays the results of the statistical analysis, numerically
showing the agreement between the data and various calculations in terms of
the $\chi^2$ per degrees of freedom and the confidence level of each
calculation. The best descriptions are found for the correlated Glauber
calculation of Ref. \cite{glauber}, and the calculation, including CT effects,
of Ref. \cite{ralston}, assuming we allow for the mentioned floating normalization.
The CT effects of the latter calculation imply only a 1\% increase in
nuclear transparency from $Q^2$ = 4 to 9 (GeV/c)$^2$,
assuming an end point-dominated distribution. Our data rule out any CT
effects larger than 7\% over the $Q^2$-range between 1.8 and 8.1 (GeV/c)$^2$,
with a confidence level of at least 90\%, but are consistent with calculations
incorporating CT effects of a few percent only, or no CT effects at all up to
$Q^2$ = 8.1 (GeV/c)$^2$.

\section{Conclusions}

Nuclear transparencies have been derived from a PWIA analysis of quasielastic
(e,e$^\prime$p) scattering from deuterium, carbon, and iron nuclei up to
$Q^2$ = 8.1 (GeV/c)$^2$.

The $A$- and $Q^2$-dependence of these nuclear transparencies were
investigated in a search for the onset of the CT phenomenon.
The $A$-dependence was parameterized as nuclear transparency
$T(Q^2) = cA^{\alpha(Q^2)}$. Using deuterium, carbon, and iron data
we find, within uncertainties, the constant $c$ to be unity as expected,
and no $Q^2$-dependence of $\alpha$, up to $Q^2$ = 8.1 (GeV/c)$^2$.
Alternatively, one can analyze the nuclear transparency data within a
simple geometric model, using the effective proton-nucleon
cross section as a free parameter \cite{sigeff}. We consistently find
an effective proton-nucleon cross section with similar energy dependence
as the free proton-nucleon cross section.
Thus, using the experimental energy dependence of the free proton-nucleon
cross section may be sufficient to describe the nuclear transparencies
we measured in a detailed Glauber calculation.

In addition, we have performed a statistical analysis of the $Q^2$ dependence
of the nuclear transparencies determined from the $^{12}$C(e,e$^\prime$p)
reaction, in comparison with state-of-the-art calculations with and
without CT effects \cite{glauber,frankfurt,nikolaev,ralston}.
Combining the $A$- and $Q^2$-dependence analysis results, we find no
evidence for the onset of CT within our range of $Q^2$.

\section{Acknowledgements}

The authors would like to thank Drs. Jain, Ralston, and Strikman for
disclosing results of their calculations prior to publication.
This work was supported by grants of the United States Department of Energy 
and the National Science Foundation.
The Southeastern Universities Research Association (SURA) operates the Thomas
Jefferson National Accelerator Facility for the United States Department of
Energy under contract DE-AC05-84ER40150.

\noindent $^a$ Present address: University of Massachusetts, Amherst, MA 01003

\noindent $^b$ Present address: Argonne National Laboratory, Argonne, IL 60439

\noindent $^c$ Present address: TJNAF, Newport News, VA 23606

\noindent $^d$ Present address: St. Mary's University, Halifax, NS, Canada B3H 3C3

\begin{table}
\squeezetable
\label{kin_tbl}
\caption{Kinematics for the present experiment. The quasi-free angles are
indicated in bold face.}
\begin{tabular}{|c|c|c|c|c|}
Average & Electron &  $Q^2$ & Electron       & Proton         \\
$T_{p^\prime}$   &  Energy  &        & $\theta_{LAB}$ & $\theta_{LAB}$ \\
(MeV)   & (GeV)    & (GeV/c)$^2$ & (degrees) & (degrees)      \\
\hline
1760 & 3.059  & 3.3 & 54.00 & 19.78,22.30,{\bf 24.81},27.28,29.78 \\
3263 & 4.463  & 6.1 & 64.65 & {\bf 15.33} \\
4293 & 5.560  & 8.1 & 64.65 & {\bf 12.84} \\ 
\end{tabular}
\end{table}

\begin{table}
\squeezetable
\label{trans_tbl}
\caption{Measured transparencies for D, C, and Fe. The first
uncertainty quoted is statistical, the second systematic. In the figures
these are added in quadrature. The uncertainties in the figures do not
include model-dependent systematic uncertainties on the simulations.
We note that a renormalization of these nuclear transparencies with a factor of
1.020 ($T_C$) and 0.896 ($T_{Fe}$) is advocated in Ref. \protect\cite{zhalov}.}
\begin{tabular}{|c|c|c|c|}
$Q^2$ & $T_D$ & $T_C$ & $T_{Fe}$ \\
(GeV/c)$^2$ & &       &          \\
\hline
3.3 & 0.897$\pm$0.013$\pm$0.027 & 0.548$\pm$0.005$\pm$0.013 & 0.394$\pm$0.009$\pm$0.009 \\
6.1 & 0.917$\pm$0.013$\pm$0.028 & 0.570$\pm$0.007$\pm$0.013 & 0.454$\pm$0.015$\pm$0.018 \\
8.1 & 0.867$\pm$0.020$\pm$0.026 & 0.573$\pm$0.010$\pm$0.013 & 0.391$\pm$0.012$\pm$0.009 \\
\end{tabular}
\end{table}

\begin{table}
\squeezetable
\label{sigeff_tbl}
\caption{Results of the fits to the A-dependence (see text) using the world's data.
Please note that the values quoted for $\sigma_{\rm eff}$ follow the framework
of Ref. \protect\cite{sigeff}, and numerical values differ slightly from those
quoted in Ref. \protect\cite{ne18a}.}
\begin{tabular}{|c|c|c|c|}
$Q^2$        & Ref. & $\alpha$ & $\sigma_{\rm eff}$\\
(GeV/c)$^2$  &      &          &     (mb)      \\
\hline
0.3 & \protect\cite{bates} (Bates) & -0.23$\pm$0.03 & 17$\pm$3 \\
0.6 & \protect\cite{donprl} (JLab) & -0.17$\pm$0.04 & 24$\pm$4 \\
1.0 & \protect\cite{ne18q,ne18a} (SLAC) & -0.18$\pm$0.02 & 22$\pm$3 \\
1.3 & \protect\cite{donprl} (JLab) & -0.22$\pm$0.05 & 27$\pm$3 \\
1.8 & \protect\cite{donprl} (JLab) & -0.24$\pm$0.04 & 32$\pm$3 \\
3.1 & \protect\cite{ne18q,ne18a} (SLAC) & -0.24$\pm$0.02 & 30$\pm$3 \\
3.3 & \protect\cite{donprl} (JLab) & -0.25$\pm$0.04 & 30$\pm$3 \\
3.3 & present work          & -0.24$\pm$0.02 & 35$\pm$3 \\
5.0 & \protect\cite{ne18q,ne18a} (SLAC) &-0.24$\pm$0.02 & 33$\pm$4 \\
6.1 & present work          & -0.24$\pm$0.03 & 30$\pm$4 \\
6.8 & \protect\cite{ne18q,ne18a} (SLAC) &-0.20$\pm$0.02 & 24$\pm$4 \\
8.1 & present work          & -0.23$\pm$0.02 & 33$\pm$3 \\
\end{tabular}
\end{table}

\begin{table}
\squeezetable
\label{ct_tbl}
\caption{Statistical comparisons of $^{12}$C(e,e$^\prime$p) data with
various model calculations. The first model is a Glauber calculation only,
the alternative models incorporate Color Transparency effects.
We also added entries assuming a floating normalization in these models,
to take into account uncertainties both in assumptions made in the Glauber
calculations and in the analysis.}
\begin{tabular}{|c|c|cc|}
Ref. & Normalization & $\chi^2$/d.f. & conf. level \\
\hline
\protect\cite{glauber} (Glauber) & Fixed    & 0.84 & 55\% \\
\protect\cite{frankfurt} (Glauber) & Fixed  & 1.82 & 9\% \\
\protect\cite{frankfurt} (+ CT)  & Fixed    & 9.4  & $<$0.1\% \\
                                 & Floating & 1.67 & 12\% \\
\protect\cite{nikolaev} (+ CT)   & Fixed    & 10.1 & $<$0.1\% \\
                                 & Floating & 1.87 & 9\% \\
\protect\cite{ralston} (+ CT)    & Fixed    & 7.8  & $<$0.1\% \\
                                 & Floating & 0.86 & 52\% \\
\end{tabular}
\end{table}

\begin{figure}
\begin{center}
\epsfxsize=3.25in
\epsfysize=4.75in
\epsffile{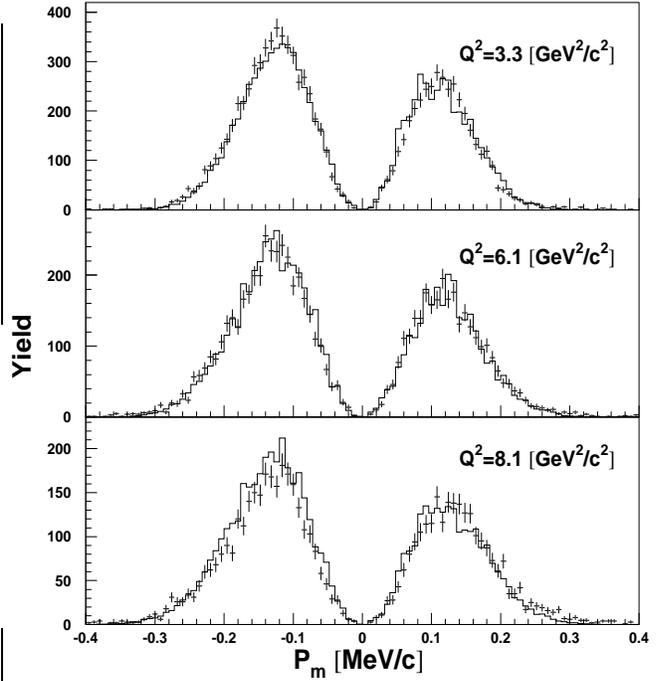}
\caption{Experimental yield (pluses) as a function of missing momentum for the 
$^{12}$C(e,e$^\prime$p) reaction, with the hadron spectrometer
positioned at the quasi-free angle, compared to simulated yields (histogram),
at $Q^2$ = 3.3, 6.1, and 8.1 (GeV/c)$^2$.
The data are integrated over a missing energy region up to 80 MeV.
Positive (negative) missing momentum is defined as a proton angle
larger (smaller) than the momentum transfer angle.}
\end{center}
\end{figure}

\begin{figure}
\begin{center}
\epsfxsize=3.25in
\epsfysize=4.75in
\epsffile{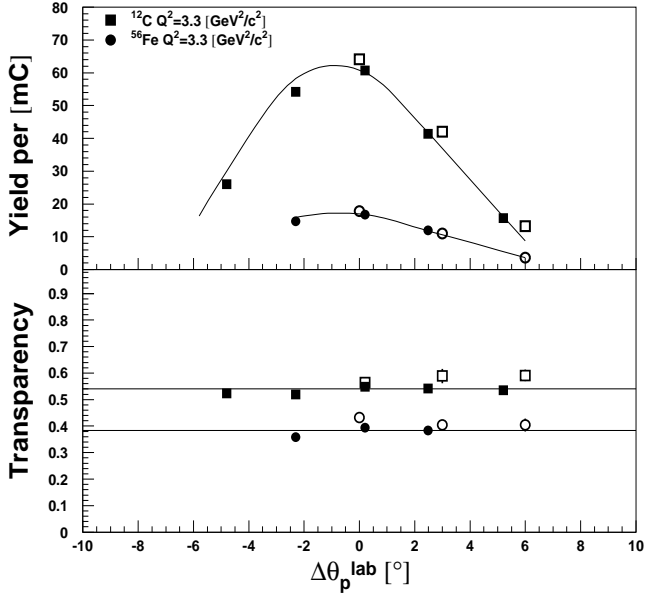}
\caption{(Upper panel) Experimental (e,e$^\prime$p) coincidence yields
vs. the difference between the proton spectrometer lab angle and the
quasi-free angle for data from $^{12}$C(e,e$^\prime$p) and
$^{56}$Fe(e,e$^\prime$p) at $Q^2$ = 3.3 (GeV/c)$^2$. Closed
symbols are for the present experiment. Open symbols are for the data from
Ref. \protect\cite{donprl}.
(Lower panel) Transparency as function of proton angle for the same data.
The curves in each panel are simulations of the yield based on the model
described in the text and normalized by a single transparency factor.}
\end{center}
\end{figure}

\begin{figure}
\begin{center}
\epsfxsize=3.25in
\epsfysize=4.75in
\epsffile{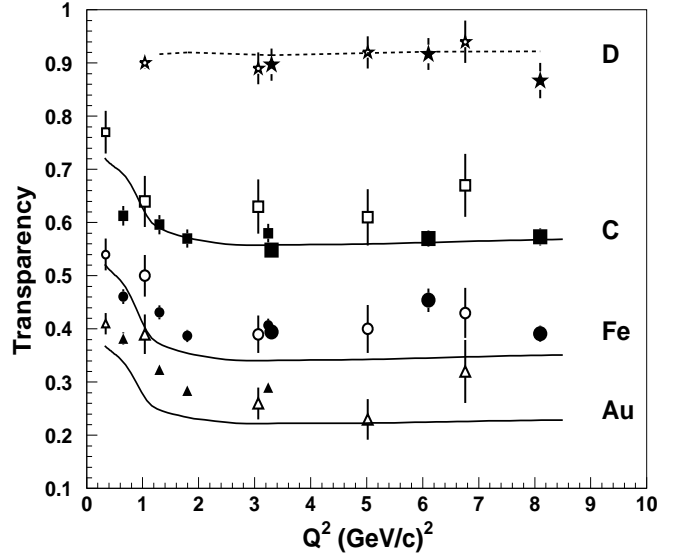}
\caption{Transparency for (e,e$^\prime$p) quasielastic scattering from D
(stars), C (squares), Fe (circles), and Au (triangles). Data from the present
work are the large solid stars, squares, and circles, respectively.
Previous JLab data (small solid squares, circles, and triangles) are from Ref.
\protect\cite{donprl}. Previous SLAC data (large open symbols) are from Ref.
\protect\cite{ne18q,ne18a}. Previous Bates data (small open symbols) at the
lowest $Q^2$ on C, Ni, and Ta targets, respectively, are from Ref.
\protect\cite{bates}. The errors shown include statistical and systematic
($\pm$ 2.3\%) uncertainties, but do not include model-dependent systematic
uncertainties on the simulations. The solid curves shown from 0.2 $<$ $Q^2$
$<$ 8.5 (GeV/c)$^2$ are Glauber calculations from Ref. \protect\cite{glauber}.
In the case of D, the dashed curve is a Glauber calculation 
from Ref. \protect\cite{sargsian}.
\label{nucl_transp}}
\end{center}
\end{figure}

\begin{figure}
\begin{center}
\epsfxsize=3.25in
\epsfysize=4.75in
\epsffile{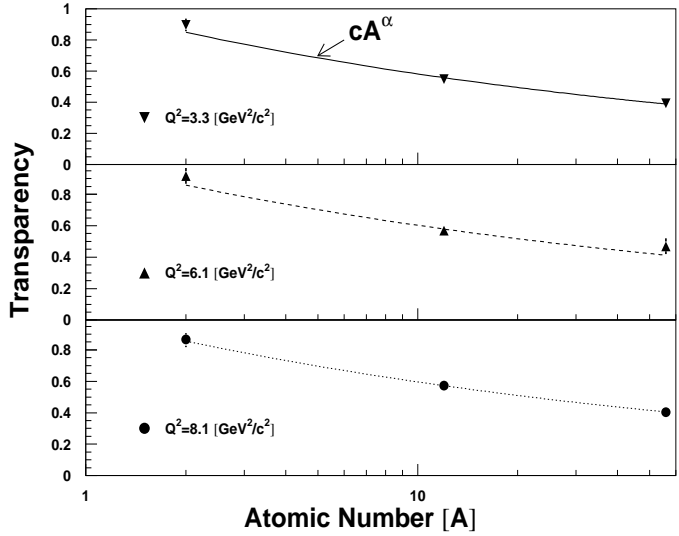}
\caption{Nuclear Transparency as a function of $A$ at $Q^2$ = 3.3, 6.1,
and 8.1 (GeV/c)$^2$ (top to bottom). The curves are fits to the D, C, and
Fe data using $T = cA^\alpha$.}
\end{center}
\end{figure}

\begin{figure}
\begin{center}
\epsfxsize=3.25in
\epsfysize=2.75in
\epsffile{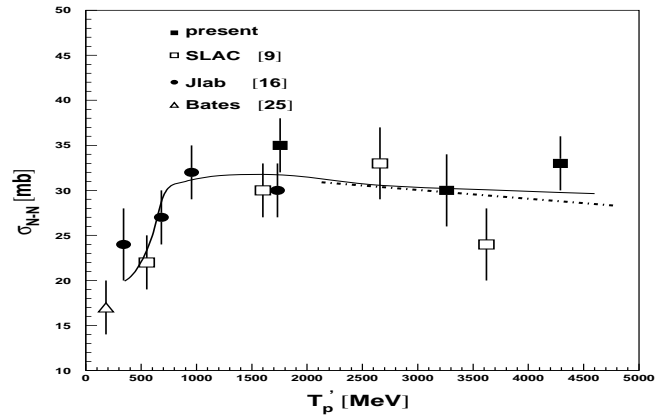}
\caption{Effective proton-nucleon cross section 
$\sigma_{\rm eff}$ as determined
using a model assuming classical attenuation of protons propagating in the
nucleus, with $\sigma_{\rm eff}$, independent of density, as fit parameter (see
text). The data are a compilation of the present work and previous work
at JLab \protect\cite{donprl}, SLAC \protect\cite{ne18a}, and Bates
\protect\cite{bates}.
The solid curve is a fit to the effective nucleon-nucleon cross sections,
assuming a similar energy dependence as the average of the free proton-proton
and proton-neutron cross sections from the Particle Data Booklet
tables \protect\cite{pdb}.
The dot-dashed curve, almost coinciding with the
solid curve, is the result of a calculation of Ref. \protect\cite{ralston}.
}
\end{center}
\end{figure}

\begin{figure}
\begin{center}
\epsfxsize=3.25in
\epsfysize=4.25in
\epsffile{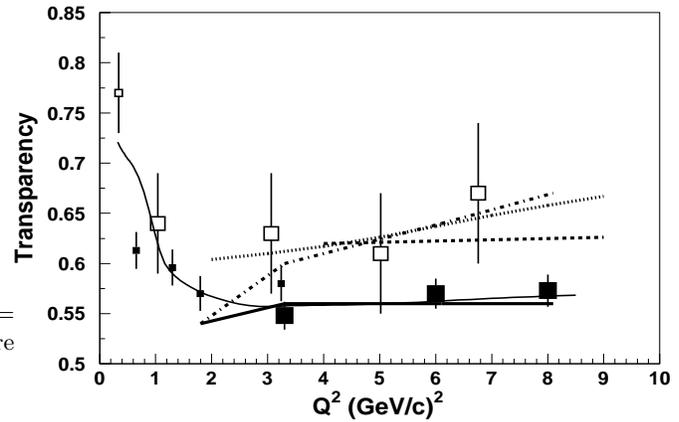}
\caption{Nuclear Transparency for $^{12}$C(e,e$^\prime$p) quasielastic
scattering. Symbols and thin solid curve are identical to
Fig.~\ref{nucl_transp}. The thick solid curve is a Glauber calculation
of Ref. \protect\cite{frankfurt}.
The dot-dashed, dotted, and dashed curves are Color Transparency predictions
from Refs. \protect\cite{frankfurt}, \protect\cite{nikolaev}, and
\protect\cite{ralston}, respectively.}
\end{center}
\end{figure}

\end{document}